\documentclass{PoS-hep}
\usepackage{amsmath}
\usepackage{bm}
\usepackage{epsfig}
\usepackage{graphics}
\usepackage{eufrak}
\usepackage{cite}
\usepackage{epsfig}
\usepackage{amssymb}
\usepackage{latexsym}
\usepackage{times}
\usepackage{slashed}
\usepackage{upgreek}
\usepackage{amsmath}
\usepackage{amsfonts}
\usepackage{amsbsy}
\usepackage{amscd}
\usepackage{bbm}

\newcommand{\eec}{\end{center}}
\newcommand{\bec}{\begin{center}}

\newcommand{\eem}{\end{matrix}}
\newcommand{\bem}{\begin{matrix}}
\newcommand{\eeq}{\end{equation}}
\newcommand{\beq}{\begin{equation}}
\newcommand{\ba}{\begin{array}}
\newcommand{\ea}{\end{array}}
\newcommand{\bea}{\begin{eqnarray}}
\newcommand{\eea}{\end{eqnarray}}
\newcommand{\baq}{\begin{eqnarray}}
\newcommand{\eaq}{\end{eqnarray}}
\newcommand{\beqs}{\begin{subequations}}
\newcommand{\eeqs}{\end{subequations}}
\newcommand{\bel}{\begin{align}}
\newcommand{\eal}{\end{align}}

\newcommand\eqs[2]{Eqs.~(\ref{#1}) and (\ref{#2})}

\newcommand\eqss[3]{Eqs.~(\ref{#1}), (\ref{#2}), and (\ref{#3})}

\newcommand{\ftn}{\footnotesize}

\newcommand{\TeV}{{\mbox{\rm TeV}}}
\newcommand{\MeV}{{\mbox{\rm MeV}}}
\newcommand{\GeV}{{\mbox{\rm GeV}}}

\newcommand{\EeV}{{\mbox{\rm EeV}}}
\newcommand{\PeV}{{\mbox{\rm PeV}}}
\newcommand{\ZeV}{{\mbox{\rm ZeV}}}
\newcommand{\YeV}{{\mbox{\rm YeV}}}
\newcommand{\ReV}{{\mbox{\rm ReV}}}

\newcommand{\etal}{{\it et al.\/}}

\def\lf{\left(}
\def\rg{\right)}
\newcommand\vev[1]{\langle {#1} \rangle}
\newcommand\vevi[1]{\langle {#1} \rangle_{\rm I}}
\newcommand{\Gr}{\ensuremath{\widetilde{G}}}

\newcommand{\Nhi}{\ensuremath{N_{\rm I\star}}}
\newcommand{\ks}{\ensuremath{k_\star}}
\newcommand{\Gsm}{\ensuremath{\mathbb{G}_{\rm SM}}}

\newcommand{\Vhi}{\ensuremath{V_{\rm I}}}
\newcommand{\vf}{\ensuremath{V_{\rm F}}}
\newcommand{\Hhi}{\ensuremath{H_{\rm I}}}
\newcommand{\Whi}{\ensuremath{W_{\rm I}}}

\newcommand{\Vhio}{\ensuremath{V_{\rm I0}}}

\newcommand{\mP}{\ensuremath{m_{\rm P}}}

\newcommand{\Ggut}{\ensuremath{\mathbb{G}}}

\newcommand{\Gfl}{\ensuremath{\mathbb{G}_{\rm 5_X}}}
\newcommand{\Glr}{\ensuremath{\mathbb{G}_{\rm LR}}}
\newcommand{\Gbl}{\ensuremath{\mathbb{G}_{B-L}}}

\newcommand{\la}{\ensuremath{\lambda}}
\newcommand{\lm}{\ensuremath{\lambda_\mu}}
\newcommand{\aS}{\ensuremath{{\rm a}_S}}

\newcommand{\Gsn}{\ensuremath{\Gamma_{\rm I}}}
\newcommand{\msn}{\ensuremath{m_{\rm I}}}
\newcommand{\mgr}{\ensuremath{m_{3/2}}}
\newcommand{\mgri}{\ensuremath{m_{\rm I3/2}}}

\newcommand{\hd}{{\ensuremath{H_d}}}
\newcommand{\hu}{{\ensuremath{H_u}}}
\newcommand{\Nr}{\ensuremath{{\sf N}_{\rm G}}}

\newcommand{\ns}{\ensuremath{n_{\rm s}}}
\newcommand{\as}{\ensuremath{\alpha_{\rm s}}}

\newcommand{\om}{\ensuremath{\omega}}

\newcommand{\Dmax}{\ensuremath{\Delta_{\rm max*}}}
\newcommand{\Dex}{\ensuremath{\Delta_{\rm c\star}}}

\newcommand{\br}{\ensuremath{{\sf B}_{\rm h}}}

\newcommand{\Trh}{\ensuremath{T_{\rm rh}}}

\newcommand{\sg}{\ensuremath{\sigma}}
\newcommand{\sgx}{\ensuremath{\sigma_\star}}
\newcommand{\sgc}{\ensuremath{\sigma_{\rm c}}}
\newcommand{\sgm}{\ensuremath{\sigma_{\rm max}}}
\newcommand{\sgf}{\ensuremath{\sigma_{\rm f}}}
\newcommand{\ld}{\ensuremath{\lambda}}

\newcommand{\kp}{\ensuremath{\kappa}}

\def\ssb{\leavevmode\hbox{$\diagup$\kern-12pt\ftn\scshape
susy}}

\newcommand{\mss}{\ensuremath{\widetilde m}}

\newcommand{\arxiv}[1]{{\ftn\tt  arXiv:#1}}

\newcommand{\fref}[1]{Fig.~\ref{#1}}

\newcommand{\Eref}[1]{Eq.~(\ref{#1})}
\newcommand{\Sref}[1]{Sec.~\ref{#1}}
\newcommand{\Fref}[1]{Fig.~\ref{#1}}
\newcommand{\Tref}[1]{Table~\ref{#1}}
\newcommand{\cref}[1]{Ref.~\cite{#1}}

\def\th{{\theta}}
\def\ths{{\theta_S}}

\def\Ka{K\"{a}hler potential}
\def\Kaa{K\"{a}hler}
\def\Kam{K\"{a}hler manifold}

\def\nano{{\sf\small NG15}}

\newcommand{\plk}{{\it Planck}}
\newcommand{\zm}{\ensuremath{Z_{-}}}

\newcommand{\bdhh}{{\ensuremath{\normalsize I{\kern-2.9pt H}}}}

\newcommand{\phc}{\ensuremath{\Phi}}
\newcommand{\phcb}{\ensuremath{\bar\Phi}}
\newcommand{\what}{\ensuremath{\widehat}}

\newcommand{\mgro}{\ensuremath{m_{3/2}}}
\newcommand{\mz}{\ensuremath{m_{z}}}
\newcommand{\mzi}{\ensuremath{m_{{\rm I}z}}}
\newcommand{\mth}{\ensuremath{m_{\theta}}}
\newcommand{\mthi}{\ensuremath{m_{\rm I\theta}}}

\newcommand{\no}{\ensuremath{N}}

\def\al{{\alpha}}

\def\bz{{Z^*}}

\newcommand{\Erefs}[2]{Eqs.~(\ref{#1}) -- (\ref{#2})}

\newcommand{\mcs}{\ensuremath{{G\mu_{\rm cs}}}}
\newcommand{\rcs}{\ensuremath{{r_{\rm cs}}}}
\newcommand{\ecs}{\ensuremath{{\epsilon_{\rm cs}}}}
\newcommand{\rms}{\ensuremath{{r_{\rm ms}}}}

\newcommand{\Ns}{\ensuremath{{N_{\rm I\star}}}}

\newcommand{\khh}{\ensuremath{K_{\rm H}}}

\newcommand{\dK}{\ensuremath{K_\mu}}

\newcommand{\dz}{\ensuremath{{\delta} z}}
\newcommand{\dzh}{\ensuremath{\what{\delta z}}}

\renewcommand{\Gsn}{\ensuremath{{\Gamma}_{\dz}}}
\newcommand{\Gth}{\ensuremath{{\Gamma}_{\th}}}
\newcommand{\Gh}{\ensuremath{{\Gamma}_{\tilde{h}}}}

\newcommand{\hh}{{\ensuremath{%\normalsize
I{\kern-2.6pt h}}}}

\renewenvironment{subequations}{%
\refstepcounter{equation}%
\setcounter{parentequation}{\value{equation}}%
  \setcounter{equation}{0}
  \ignorespaces
}{%
  \setcounter{equation}{\value{parentequation}}%
  \ignorespacesafterend
}

\title{\boldmath \bfseries F-Term Hybrid Inflation and  SUSY Breaking}

\ShortTitle{FHI and SUSY Breaking}

\author{\speaker{C. Pallis}\\
School of Civil Engineering, \\ Faculty of Engineering, \\
Aristotle University of Thessaloniki, \\ GR-541 24 Thessaloniki, GREECE \\
E-mail: \email{kpallis@auth.gr}}

\abstract{We consider F-term hybrid inflation and supersymmetry
breaking in the context of a model which largely respects a global
$U(1)$ $R$ symmetry. The K\"ahler potential parameterizes the
K\"ahler manifold with an enhanced $U(1)_R\times(SU(1,1)/U(1))$
symmetry, where the scalar curvature of the second factor is
determined by the achievement of a supersymmetry-breaking de
Sitter vacuum without ugly tuning. The magnitude of the emergent
soft tadpole term for the inflaton can be adjusted in the range
$(1.2-460)~\TeV$ -- increasing with the dimensionality of the
representation of the waterfall fields -- so that the inflationary
observables are in agreement with the observational requirements.
The mass scale of the supersymmetric partners turns out to lie in
the region $(0.09-253)~\PeV$ which is compatible with high-scale
supersymmetry and the results of LHC on the Higgs boson mass. The
$\mu$ parameter can be generated by conveniently applying the
Giudice-Masiero mechanism and assures the out-of-equilibrium decay
of the $R$ saxion at a low reheat temperature $\Trh\leq163~\GeV$.

}

\FullConference{Corfu Summer Institute 2024 "School and Workshops
on Elementary Particle Physics and Gravity" (CORFU2024)
12 - 26 May and 25 August - 27 September, 2024\\
Corfu, Greece}

\begin{document}

\section{Motivation}

The \emph{Supersymmetric} ({\sf\ftn SUSY}) hybrid inflation
\cite{hisusy} based on F terms, called for short henceforth
\emph{F-term hybrid inflation} ({\sf\ftn FHI}) is undoubtedly a
well-motivated, flexible and easily embedded in Particle models
inflationary model for a number of reasons. Most notably:

\begin{itemize}

\item It is based on a renormalizable superpotential uniquely
determined by a gauge  $\Ggut$ and a global $U(1)~R$ symmetries;

\item  It does not require fine tuned parameters and
transplanckian inflaton values;

\item It can be naturally followed by a \emph{Grand Unified
Theory} ({\sf \ftn GUT}) phase transition which may lead to the
production of cosmological defects, if these are predicted by the
symmetry-breaking scheme.

\end{itemize}

However, the original version of FHI \cite{hisusy}, which employs
only \emph{radiative corrections} ({\sf \ftn RCs}) into the
inflationary potential is considered as strongly disfavored by the
\plk\ data \cite{plin} due to the large scalar spectral index.
This conclusion can be evaded, if we take in to account soft
SUSY-breaking terms \cite{sstad,mfhi,kaihi} and
\emph{Supergravity} ({\sf \ftn SUGRA}) corrections \cite{hinova}
with appropriate magnitude. Both corrections above are related to
the adopted SUSY-breaking or \emph{hidden sector} ({\ftn\sf HS})
sector of the theory. Therefore, it is crucial to obtain a
consistent interconnection of HS with the \emph{inflationary
sector} ({\ftn\sf IS}) of FHI.

A possible combination of the aforementioned sectors is presented
in \Sref{asfhi} following \cref{asfhi} -- see also \cref{blfhi}.
Then, in \Sref{fhi} we reconcile the inflationary observables with
data and describe the post-inflationary evolution in \Sref{pres}
which is related to the \emph{Dark Energy} ({\ftn\sf DE}) problem,
the immunity of \emph{Big Bang Nucleosynthesis} ({\sf\ftn BBN})
from the notorious moduli problem and the \emph{Gravitational
Waves} ({\sf\ftn GWs}) obtained from the decay of \emph{Cosmic
Strings} ({\sf\ftn CSs}). Our conclusions are summarized in
\Sref{con}.

\section{Linking FHI With a SUSY-Breaking Sector}\label{asfhi}

To achieve a consistent combination of HS and IS we use the $R$
symmetry, which is a crucial ingredient for the implementation of
FHI, as a junction mechanism. Below we specify the particle
content, the superpotential, and the \Ka\ of our model in
Secs.~\ref{asfhi1aa}, \ref{md1} and \ref{md2} respectively.

\subsection{Particle Content}\label{asfhi1aa}

FHI can be implemented by introducing three superfields
$\bar{\Phi}$, $\Phi$ and $S$. The two first are left-handed chiral
superfields oppositely charged under a gauge group $\Ggut$ whereas
the latter is the inflaton and is a $\Ggut$-singlet left-handed
chiral superfield. Singlet under $\Ggut$ is also the SUSY breaking
(goldstino) superfield $Z$. In this work we identify $\Ggut$ with
three possible gauge groups with different dimensionalities $\Nr$
of the representations to which $\bar{\Phi}$ and $\Phi$ belong --
see \Tref{tab1}. Namely, we consider the following $\Ggut$'s
\beqs\bea \label{gbl} &\Gbl:= \Gsm\times U(1)_{B-L}\hspace*{3.4cm} &~~\mbox{with}~~\Nr=1,\\
\label{glr} &\Glr:= SU(3)_{\rm C}\times SU(2)_{\rm L} \times
SU(2)_{\rm R} \times U(1)_{B-L} &~~\mbox{with}~~\Nr=2,\\
\label{gfl} &\Gfl:= SU(5)\times U(1)_X \hspace*{3.7cm}
&~~\mbox{with}~~\Nr=10. \eea
%
%\beqs\bea \label{gbl} \Gbl&:=& \Gsm\times U(1)_{B-L}\\
%\label{glr} \Glr&:=& SU(3)_{\rm C}\times SU(2)_{\rm L} \times
%SU(2)_{\rm R} \times U(1)_{B-L}\\
%\label{gfl} \Gfl&:=& SU(5)\times U(1)_X. \eea\eeqs
%
Here \Gsm\ is the well-known gauge group of the SM
\beq\label{gsm} \Gsm:= SU(3)_{\rm C}\times SU(2)_{\rm L} \times
U(1)_{Y},\eeq\eeqs
to which $\Ggut$ is broken via the \emph{vacuum expectation
values} ({\ftn\sf v.e.vs}) of $\Phi$ and $\bar\Phi$ at the end of
FHI. As regards the cosmological defects, CSs are produced only
for $\Ggut=\Gbl$ -- see \Sref{cssec}.

\renewcommand{\arraystretch}{1.3}
\begin{table}[t] \bec\begin{tabular}{|c|c|c|c|c|}\hline
{\sc Super-} &\multicolumn{3}{c|}{\sc Representations Under
$\Ggut$}&$R$\\\cline{2-4}
{\sc Fields}&\Gbl&\Glr&\Gfl&{\sc Charge}\\\hline\hline
\multicolumn{5}{|c|}{\sc Higgs Superfields}\\\hline
$\phc$&$({\bf 1, 1}, 0, 2)$&$({\bf 1, 1, 2}, 1)$&$({\bf 10}, 1)$&0\\
$\phcb$&$({\bf 1, 1}, 0, -2)$&$({\bf 1, 1, \bar 2}, -1)$&$({\bf \overline{10}}, -1)$&0\\
$S$&$({\bf 1, 1}, 0, 0)$&$({\bf 1, 1, 1}, 0)$&${\bf 1}$&$2$\\
\hline
\multicolumn{5}{|c|}{\sc Goldstino Superfield}\\\hline
$Z$&$({\bf 1, 1}, 0, 0)$&$({\bf 1, 1, 1}, 0)$&${\bf
1}$&$2/\nu$\\\hline
\end{tabular}\eec
\caption{\sl Representations and $R$ charges of the superfields
involved in the IS and HS for the various $\Ggut$'s.}\label{tab1}
\end{table}
\renewcommand{\arraystretch}{1.}

\subsection{Superpotential}\label{md1}

The superpotential of our model carries $R$ charge 2 and is linear
\emph{with respect to} ({\ftn\sf w.r.t.}) $S$ and $Z^{\nu}$. It
naturally splits into four parts:
\beq \label{Who} W=W_{\rm I} +W_{\rm H} +W_{\rm GH} +W_{\rm
Y},\eeq
where the subscripts ``I'' and ``H'' stand for the IS and HS
respectively and the content of each term is specified as follows:

\subparagraph{\sf\ftn (a)} $\Whi$ is the part of $W$ related to IS
\cite{hisusy}:
\beqs\beq \Whi = \kp S\left(\bar
\Phi\Phi-M^2\right),\label{whi}\eeq
where $\kp$ and $M$ are free parameters which may be constrained
by the inflationary requirements -- see \Sref{fhi4}.

\subparagraph{\sf\ftn (b)} $W_{\rm H}$ is the part of $W$ devoted
to HS \cite{susyr}:
\beq W_{\rm H} = m\mP^2 (Z/\mP)^\nu, \label{wh} \eeq
where $\mP=2.4~\ReV$ is the reduced Planck mass -- with
$\ReV=10^{18}~\GeV$ --, $m$ is a positive free parameter with mass
dimensions, and $\nu$ is an exponent which may, in principle,
acquire any real value if $W_{\rm H}$ is considered as an
effective superpotential valid close to non-zero $\vev{Z}$. We
assume though that $\nu>0$. If we also assume that $W$ is
holomorphic in $S$ then mixed terms of the form
$S^{\nu_s}Z^{\nu_z}$ can be forbidden in $W$ since the exponent of
a such term has to obey the relation
\bea
\nu_s+\nu_z/\nu=1~~\Rightarrow~~\nu_z=(1-\nu_s)\nu,\nonumber\eea
leading to negative values of $\nu_z$. This conclusion contradicts
with our assumptions above.

%As we see below, we confine ourselves in the range $3/4<\nu<1$.

\subparagraph{\sf\ftn  (c)} $W_{\rm GH}$ is a term which mixes $Z$
and $\phcb-\phc$
\beq W_{\rm GH} = -\la\mP(Z/\mP)^\nu \phcb\phc \label{wgh}
\eeq\eeqs
with $\la$ a real coupling constant. The magnitude of $\la$ can be
restricted by the DE requirement as we see in \Sref{des} below.

\subparagraph{\sf\ftn  (d)} $W_{\rm Y}$ includes the usual
trilinear terms with Yukawa couplings between the various
superfields of \emph{Minimal SUSY Standard Model} ({\sf\ftn
MSSM}), denoted by $Y_\al$ with $\al=1,...,7$, i.e.,
\bea Y_\al= {Q}, {L}, {d}^c, {u}^c, {e}^c,
\hd,~~\mbox{and}~~\hu,\nonumber \eea
where the generation indices are suppressed. Note, however, that
the $R$ assignments as shown in \Tref{tab2} for the electroweak
Higgs superfields prohibit the presence in $W$ of a bilinear term
of the form
\beq \label{hb} H_{\rm B} =\begin{cases}\hu\hd~&\mbox{for}~~\Ggut=\Gbl,\\\hh^2&\mbox{for}~~\Ggut=\Glr,\\
{\bf\bar 5}_h {\bf 5}_h&\mbox{for}~~\Ggut=\Gfl.\end{cases}\eeq
and other unwanted mixing terms -- e.g. $\lm S\hu\hd$, $\ld
Z^\nu\hu\hd/\mP^\nu$. We do not address the issue of the
generation of tiny neutrino masses here -- see \cref{blfhi}.

\renewcommand{\arraystretch}{1.3}
\begin{table}[t] \bec\begin{tabular}{|c|c|c|c|c|}\hline
{\sc Super-} &\multicolumn{3}{c|}{\sc Representations Under
$\Ggut$}&$R$\\\cline{2-4}
{\sc Fields}&\Gbl&\Glr&\Gfl&{\sc Charge}\\\hline\hline
$\hu$&$({\bf 1, 2}, 1/2, 0)$&&&2\\
$\hd$&$({\bf 1, 2}, -1/2, 0)$&&&2\\\hline
$\hh$&&$({\bf 1,2, 2}, 0)$&&$2$\\ \hline
${\bf 5}_h$&&&$({\bf 5},2)$&$2$\\
${\bf\bar 5}_h$&&&$({\bf\bar 5},-2)$&$2$\\
\hline
\end{tabular}\eec
\caption{\sl Representations and $R$ charges of the electroweak
Higgs superfields for the various $\Ggut$'s.}\label{tab2}
\end{table}
\renewcommand{\arraystretch}{1.}

\subsection{K\"{a}hler Potential}\label{md2}

The \Ka\ respects the adopted $\Ggut$ symmetry in \Tref{tab1}. It
has the following contributions
\beq \label{Kho} K=K_{\rm I}+K_{\rm H}+\dK+|Y_\al|^2,\eeq
which can be specified as follows:

\subparagraph{\sf\ftn (a)} $K_{\rm I}$ depends on the fields
involved in FHI -- cf. \Eref{whi}. We adopt the simplest possible
choice \cite{mfhi, hinova} that has the form
\beqs\beq K_{\rm I} = |S|^2+|\Phi|^2+|\bar\Phi|^2. \\
\label{ki} \eeq
Higher order terms of the form $|S|^{2\nu_S}/\mP^{2\nu_S-2}$ with
$\nu_S>1$ can not be excluded by the imposed symmetries but may
become harmless if $S\ll\mP$ and assume low enough coefficients.

\subparagraph{\sf\ftn (b)} $K_{\rm H}$ is devoted to HS. We adopt
the form introduced in \cref{susyr} where
\beq K_{\rm
H}=\no\mP^2\ln\lf1+\frac{|Z|^2-k^2\zm^4/\mP^2}{\no\mP^2}\rg
~~\mbox{with}~~Z_{\pm}=Z\pm Z^*. \label{khi} \eeq\eeqs
Here, $k>0$ mildly violates $R$ symmetry endowing $R$ axion with
phenomenologically acceptable mass. The selected $K_{\rm H}$ is
largely respects the $R$ symmetry, which is a crucial ingredient
for FHI, and it ensures -- as we see in \Sref{des} -- a dS vacuum
of the whole field system with tunable cosmological constant for
\beq
\no=\frac{4\nu^2}{3-4\nu}~~\mbox{with}~~\frac34<\nu<\frac32~~\mbox{for}~~\no<0.\label{no}
\eeq
Our favored $\nu$ range finally is $3/4<\nu<1$. Since $\no<0$,
$K_{\rm H}$ parameterizes the $SU(1,1)/U(1)$ hyperbolic \Kam\ for
$k\sim0$.

\subparagraph{\sf\ftn (c)} $\dK$ includes higher order terms which
generate the needed mixing term between $\hu$ and $\hd$ in the
lagrangian of MSSM \cite{susyr} and has the form
\beq \dK=\lm\lf{\bz^{2\nu}}/{\mP^{2\nu}}\rg H_{\rm B} +\ {\rm
h.c.},\label{dK}\eeq
where the dimensionless constant $\lm$ is taken real for
simplicity.

\subparagraph{} The total $K$ in \Eref{Kho} enjoys an enhanced
symmetry for the $\bar Y_A, S$ and $Z$ fields with $\bar
Y_A=Y_\al, \phcb, \phc, S$. Namely,
\beq  \prod_A U(1)_{\bar Y_A} \times  \lf SU(1,1)/U(1)\rg_Z,\eeq
where the indices indicate the moduli which parameterize the
corresponding manifolds. Thanks to this symmetry, mixing terms of
the form $S^{\tilde \nu_s}Z^{*\tilde \nu_z}$ can be ignored
although they may be allowed by the $R$ symmetry for
$\tilde\nu_z=\nu\tilde\nu_s$. Most notably, $U(1)_S$ protects
$K_{\rm I}$ from $S$ depended terms which violates the $R$
symmetry, thereby, spoiling the inflationary set-up.

%\newpage
\section{Inflation Analysis}\label{fhi}

It is well known \cite{hisusy, hinova} that in global SUSY FHI
takes place for $|S|\gg M$ along a F- and D- flat direction of the
SUSY potential
\begin{equation} \label{v0}\bar\Phi={\Phi}=0,~~\mbox{where}~~ V_{\rm SUSY}\lf{\Phi}=0\rg:=V_{\rm I0}=\kp^2
M^4~~\mbox{and}~~\Hhi=\sqrt{\Vhio/3\mP^2}\eeq
are the constant potential energy density and the correspoding
Hubble parameter which drive FHI -- the subscript $0$ means that
this is the tree level value. The same configuration can be used
as an inflationary track \cite{asfhi} if we consider the SUGRA
F-term potential $V_{\rm F}$ and stabilize the HS of the model --
see Sec.~\ref{fhi1}. Then, in \Sref{fhi2}, we give the form of the
inflationary potential and employ it to obtain our results
presented in \Sref{fhi4} after imposing a number of constraints
listed in \Sref{fhi3}.

\subsection{Hidden Sector's Stabilization}\label{fhi1}

\begin{table} \bec \begin{tabular}{|c|c|c|c||c|}\hline
%
%\multicolumn{5}{|c|}{\sc\bfseries Model Parameters}\\\hline
%
$\Ggut$&$\ld/10^{-12}$&$M/\YeV$&$m/\PeV$&$\as/\TeV$\\\hline\hline
\Gbl&$0.2$&$1.4$&$0.5$&$2.63$\\\hline
\Glr&$1.7$&$1.9$&$1.15$&$6.7$\\\hline
\Gfl&$2.6$&$3.6$&$6.3$&$56.3$\\\hline
\end{tabular}\eec
\caption{\sl Parameters of interest for the various $\Ggut$'s with
fixed $\kp=5\cdot10^{-4}$, $\nu=7/8$ and $k=0.1$ resulting to
$\ns=0.967$ -- recall that $1~\YeV=10^{15}~\GeV$.}\label{tab3}
\end{table}
\renewcommand{\arraystretch}{1.}

FHI may be attained if $Z$ is well stabilized during it to a value
well below $\mP$. To determine this, we analyze the relevant
fields as follows
\beq Z=(z+i\theta)/\sqrt{2}~~\mbox{and}~~S=\sg\
e^{i\ths/\mP}/\sqrt{2}\label{Zpara}\eeq
and we construct \cite{asfhi} the complete expression for $\vf$
along the inflationary trajectory in \Eref{v0}. Upon expanding the
resulting expression for low $S/\mP$ values and keeping $\theta=0$
as in its present vacuum -- see \Sref{des} below -- we find that
$\vf$ is minimized for the value \cite{asfhi}
\beq \label{veviz}
\vevi{z}\simeq\lf\sqrt{3}\cdot2^{\nu/2-1}\Hhi/m\nu\sqrt{1-\nu}\rg^{1/(\nu-2)}\mP.\eeq
Note that $\nu<1$ assures a real value of $\vevi{z}$ with
$\vevi{z}\ll\mP$ since $\Hhi/m\ll1$. In particular, for the inputs
of \Tref{tab3} we obtain $\vevi{z}/10^{-3}\mP=1.1, 1.5$ and $2.5$
for $\Ggut=\Gbl, \Glr$ and $\Gfl$ respectively. Furthermore,
$\vevi{z}$ turns out to be independent from $\sg$ as highlighted
in \fref{fig1} where the quantity $10^5(\vf/\kp^2M^4-1)$ is
plotted as a function of $z$ and $\sg$ with fixed $\th=0$ for the
inputs of \Tref{tab3} with $\Ggut=\Glr$. We also depict by a thick
black point the location of $(\vevi{z},\sgx)$, where $\sgx$ is the
value of $\sg$ when the pivot scale $\ks=0.05/{\rm Mpc}$ crosses
outside the horizon of FHI -- see \Sref{fhi4} below.

The (canonically normalized) components of sgoldstino, acquire
masses squared, respectively,
\beqs\beq\mzi^2\simeq6(2-\nu)\Hhi^2~~\mbox{and}~~
\mthi^2\simeq3\Hhi^2-
m^2\lf8\nu^2\mP^2-3\vevi{z}^2\rg\frac{4\nu(1-\nu)\mP^2+(1-96k^2\nu)\vevi{z}^2}{2^{3+\nu}\nu\mP^{2\nu}\vevi{z}^{2(2-\nu)}},
\label{mz8i} \eeq
whereas the mass of $\Gr$ turns out to be
\beq \mgri\simeq \lf
\nu(1-\nu)^{1/2}m^{2/\nu}/\sqrt{3}\Hhi\rg^{\nu/(2-\nu)}.\label{mgri}\eeq\eeqs
Given that $\Hhi\sim\EeV$ and taking into account the results
arranged in the Table of \fref{fig1} we can infer that
$\mzi\gg\Hhi$ and so $z$ is well stabilized during FHI whereas
$m_{\rm I\theta}\simeq\Hhi$ and gets slightly increased as $k$
increases beyond $0.1$. Therefore, no problem with the
isocurvature perturbation is expected.

\begin{figure}[t]\vspace*{-.6cm}
\begin{minipage}{75mm}
\vspace*{-4.cm}\epsfig{file=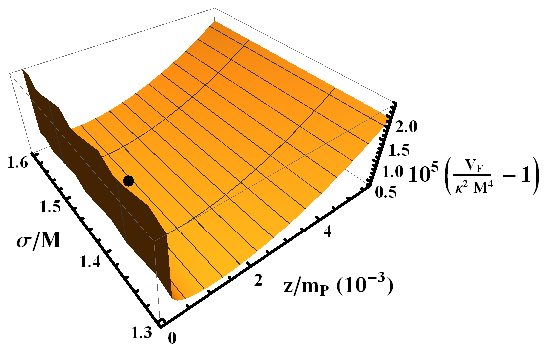,width=8.3cm,angle=-0}\end{minipage}
\begin{minipage}{75mm} \begin{center}\renewcommand{\arraystretch}{1.3}
\vspace*{.7in}\hspace*{0.4in}\begin{tabular}{|c||c|c|c|}\hline
\Ggut&$\mzi/\EeV$&$\mthi/\EeV$&$\mgri/\EeV$\\\hline\hline
\Gbl&$0.64$&$0.15$&$1.2$\\\hline
\Glr&$1.1$&$0.32$&$2.98$\\\hline
\Gfl&$4.1$&$1.2$&$25$\\\hline
\end{tabular}\eec
\end{minipage}
\vspace*{.55in}\caption{\sl The SUGRA potential
$10^5(\vf/\kp^2M^4-1)$ along the path in Eq.~(3.1) as a function
of $z$ and $\sg$ for $\th=0$ and the parameters of Table 3 for
$\Ggut=\Glr$. The location of $(\vevi{z},\sgx)$ is also depicted
by a thick black point. The particle mass spectrum during FHI for
the various $\Ggut$'s are given in the Table -- recall that
$1~\EeV=10^9~\GeV$.} \label{fig1}\end{figure}
\renewcommand{\arraystretch}{1.}

\subsection{Inflationary Potential}\label{fhi2}

Expanding $\vf$ for low $S$ values, introducing the canonically
normalized inflaton $\sg=\sqrt{2}|S|$ and taking into account the
RCs \cite{hisusy} we derive \cite{asfhi} the inflationary
potential $V_{\rm I}$ which may take the form
\beq\label{vol} V_{\rm I}\simeq V_{\rm I0}\left(1+C_{\rm
RC}+C_{\rm SSB}+C_{\rm SUGRA}\right).\eeq
The individual contributions are specified as follows:

\subparagraph{\sf\ftn (a)} $C_{\rm RC}$ represents the RCs to
$V_{\rm I}/V_{\rm I0}$ which may be written as \cite{hisusy,
hinova}
\beqs\beq \label{crc}C_{\rm RC}=
%\begin{array}{rl}
{\kappa^2\Nr\over 128\pi^2}\lf8\ln{\kp^2 M^2\over Q^2}+f_{\rm
RC}\lf\sigma\over M\rg\rg~~\mbox{with}~~\Nr = \begin{cases}1~~&\mbox{for}~~\Ggut=\Gbl,\\
2&\mbox{for}~~\Ggut=\Glr,\\ 10&\mbox{for}~~\Ggut=\Gfl.\end{cases}
\eeq
Also the function $f_{\rm RC}$ may be expressed as
\beq f_{\rm RC}(x)=8x^2\tanh^{-1}\lf{2}/{x^2}\rg-4(\ln4-x^4\ln
x)+(4+x^4)\ln(x^4-4)~~\mbox{with}~~x>\sqrt{2}.\eeq
When $\sg$ tends to a critical value $\sgc=\sqrt{2}M$, the term
including $\tanh^{-1}$ tends to infinity. This means that one
effective mass of the particle spectrum becomes negative, causing
a destabilization of $\phcb$ and $\phc$ from zero in \Eref{v0}
triggering, thereby, the $\Ggut$ phase transition.

\subparagraph{\sf\ftn (b)} $C_{\rm SSB}$ is the contribution to
$V_{\rm I}/V_{\rm I0}$ from the soft SUSY-breaking effects
\cite{sstad} parameterized as follows:
\beq \label{cssb}C_{\rm SSB}=
%\begin{array}{rl}
m_{\rm I3/2}^2 \sg^2/2V_{\rm I0}-{\rm a}_S\,\sigma /\sqrt{2V_{\rm
I0}},\eeq
where the tadpole parameter reads \cite{asfhi}
\beq \label{aSn} {\rm
a}_S=2^{1-\nu/2}m\frac{\vevi{z}^\nu}{\mP^\nu}\lf1+\frac{\vevi{z}^2}{2N\mP^2}\rg
\lf2-\nu-\frac{3\vevi{z}^2}{8\nu\mP^2}\rg.\eeq
The minus sign results from the stabilization of $\theta$ at zero
and the minimization of the factor
$(S+S^*)=\sqrt{2}\sg\cos(\theta_S/\mP)$ which occurs for
$\theta_S/\mP=\pi~({\sf mod}~2\pi)$ -- the decomposition of $S$ is
shown in \Eref{Zpara}. We further assume that $\theta_S$ remains
constant during FHI so that the simple one-field slow-roll
approximation is valid -- cf.~\cref{kaihi}.

\subparagraph{\sf\ftn (c)} $C_{\rm SUGRA}$ is the SUGRA correction
to $V_{\rm I}/V_{\rm I0}$, after subtracting the one in $C_{\rm
SSB}$, which is \cite{asfhi}
\beq \label{csugra} C_{\rm
SUGRA}=c_{2\nu}\frac{\sg^2}{2\mP^2}+c_{4\nu}\frac{\sg^4}{4\mP^4}~~\mbox{with}~~
 c_{2\nu}=\frac{\vevi{z}^2}{2\mP^2}~~\mbox{and}~~c_{4\nu}=\frac12\lf1+\frac{\vevi{z}^2}{\mP^2}\rg.
\eeq\eeqs
The establishment of FHI is facilitated by the smallness of the
coefficients above which is due to the minimality of $K_{\rm I}$
in \Eref{ki} and the stabilization of $z$ at $\vevi{z}\ll\mP$.

\subsection{Observational Requirements}\label{fhi3}

Our model of FHI can be qualified if we test it against a number
of observational requirements. Namely:

\subparagraph{\sf\ftn (a)} The number of e-foldings that the pivot
scale $\ks=0.05/{\rm Mpc}$ suffered during FHI have to be enough
to resolve the problems of the Standard Big Bang, i.e.,
\cite{plin}:
\begin{equation}  \label{Nhi}
\Ns=\int_{\sigma_{\rm f}}^{\sigma_{\star}} \frac{d\sigma}{m^2_{\rm
P}}\: \frac{V_{\rm I}}{V'_{\rm I}}\simeq19.4+{2\over
3}\ln{V^{1/4}_{\rm I0}\over{1~{\rm GeV}}}+ {1\over3}\ln {T_{\rm
rh}\over{1~{\rm GeV}}},
\end{equation}
where the prime denotes derivation w.r.t. $\sigma$, $\sgx$ is the
value of $\sigma$ when $\ks$ crosses outside the horizon of FHI
and $\sigma_{\rm f}\simeq\sgc$ signals the termination of FHI.
Also we take for the reheating temperature $T_{\rm rh}$  a value
close to those met in our set-up -- see \Sref{reh} below --
$\Trh\simeq1~\GeV$. In all cases, we obtain $\Nhi\simeq40$.

\subparagraph{\sf\ftn (b)} The amplitude $A_{\rm s}$ of the power
spectrum of the curvature perturbation generated by $\sigma$
during FHI must be appropriately normalized \cite{plcp}, i.e.,
\begin{equation} \label{Prob}
A_{\rm s}= \frac{1}{12\, \pi^2 m^6_{\rm P}}\; \left.\frac{V_{\rm
I}^{3}(\sigma_\star)}{|V'_{\rm I}(\sigma_\star)|^2}\right.\simeq\:
2.105\cdot 10^{-9}.
\end{equation}

\subparagraph{\sf\ftn (c)} The remaining observables -- the scalar
spectral index $\ns$, its running $\as$, and the scalar-to-tensor
ratio $r$ -- which are calculated by the following standard
formulas
\beq \label{ns}  \ns=1-6\epsilon_\star\ +\ 2\eta_\star,
\as={2}\left(4\eta_\star^2-(\ns-1)^2\right)/3-2\xi_\star~~\mbox{and}~~
r=16\epsilon_\star,\eeq
(where $\xi\simeq m_{\rm P}^4~V'_{\rm I} V'''_{\rm I}/V^2_{\rm I}$
and all the variables with the subscript $\star$ are evaluated at
$\sigma=\sgx$) must be in agreement with data, i.e., \cite{plin,
gws}
\beq \label{nswmap}
\ns=0.967\pm0.0074~~\mbox{and}~~r\lesssim0.032, \eeq at 95$\%$
\emph{confidence level} ({\sf\ftn c.l.}) with negligible
$|\as|\ll0.01$.

\begin{figure}[t]\vspace*{-.90cm}
\begin{minipage}{75mm}
\vspace*{.5cm}
\epsfig{file=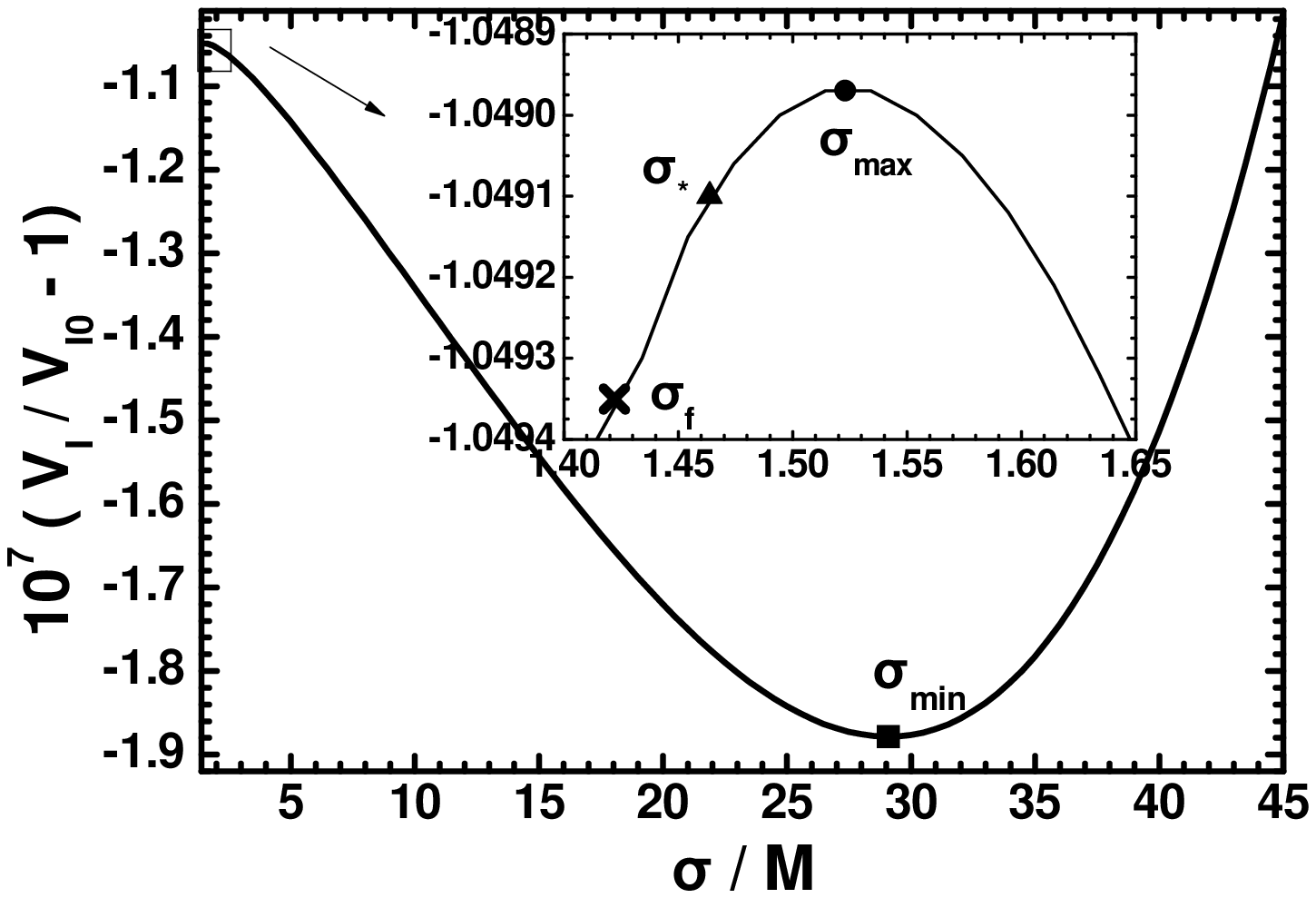,width=5.3cm,angle=-90}\end{minipage}
\begin{minipage}{75mm} \begin{center}\renewcommand{\arraystretch}{1.3}
\vspace*{.0in}\hspace*{0.4in}\begin{tabular}{|c||c|c|c|}\hline
\Ggut&$\sgx/\sqrt{2}M$&$\Dex~(\%)$&$\Dmax~(\%)$\\\hline\hline
\Gbl&$1.026$&$2.6$&$2.9$\\\hline
\Glr&$1.035$&$3.5$&$3.9$\\\hline
\Gfl&$1.067$&$6.7$&$7.3$\\\hline
\end{tabular}\eec
\end{minipage}
\vspace*{.in}\caption{\sl The variation of $\Vhi$ as a function of
$\sg$ for the parameters given in Table~3 for $\Ggut=\Glr$. The
values $\sgx$, $\sg_{\rm f}$, $\sg_{\rm max}$, and $\sg_{\rm min}$
of $\sigma$ are also depicted. Values of $\sgx, \Dex$ and $\Dmax$
for the inputs of Table~3 are also given in the Table for the
various $\Ggut$'s.} \label{fig3}\end{figure}

\subsection{Results}\label{fhi4}

%We here present the $\aS$ values as a function of $\kp$ or $M$
%which assist the confrontation of FHI with data and postpone its
%derivation from the fundamental parameters $m$ and $\nu$ via
%\Eref{aSn} in \Sref{susy}.

Enforcing  Eqs.~(\ref{Nhi}) and (\ref{Prob}) we can restrict $M$
and $\sgx$ as functions of $\kappa$ for any given $\aS$. The
correct values of $\ns$ in \Eref{nswmap} are attained if we
carefully select $\aS$ so that $\Vhi$ in \Eref{vol} becomes
non-monotonic and develops a maximum at $\sg=\sgm$ and a minimum
at $\sg_{\rm min}\gg\sgm$ as shown in \Fref{fig3}. The relevant
plot visualizes the variation of $\Vhi$ as a function of $\sg$ for
the parameters given in Table~3 with $\Ggut=\Glr$. From the
subplot of this figure we remark that $\sgf<\sgx<\sgm$ as suited
for hilltop FHI \cite{mfhi, kaihi}. To qualify the relevant tuning
we define the quantities
\beq \Dex={(\sgx-\sgc)/\sgc}~~\mbox{and}~~\Dmax=
{(\sgm-\sgx)/\sgm}\,.\label{dms}\eeq
The naturalness of the hilltop FHI increases with $\Dex$ and
$\Dmax$. To get an impression of the amount of these tunings and
their dependence on the parameters of the model, we display in the
Table of \Fref{fig3}  the resulting $\Dex$ and $\Dmax$ together
with $\sgx$ for the values of \Tref{tab3}. We notice that
$\Dmax>\Dex$ and that their values may be up to $10\%$ increasing
with $\Nr$ (and $\aS$). From our data we infer that
$|\as|\sim10^{-4}$ and $r\sim10^{-11}$, and so beyond the reach of
the planned experiments aiming to detect primordial GWs. For the
$\ns$ values in \Eref{nswmap}, we observe that $r$ and $|\as|$
increase with $\aS$.

\begin{figure}[!t]\vspace*{-.17in}
\hspace*{-.12in}
\begin{minipage}{8in}
\epsfig{file=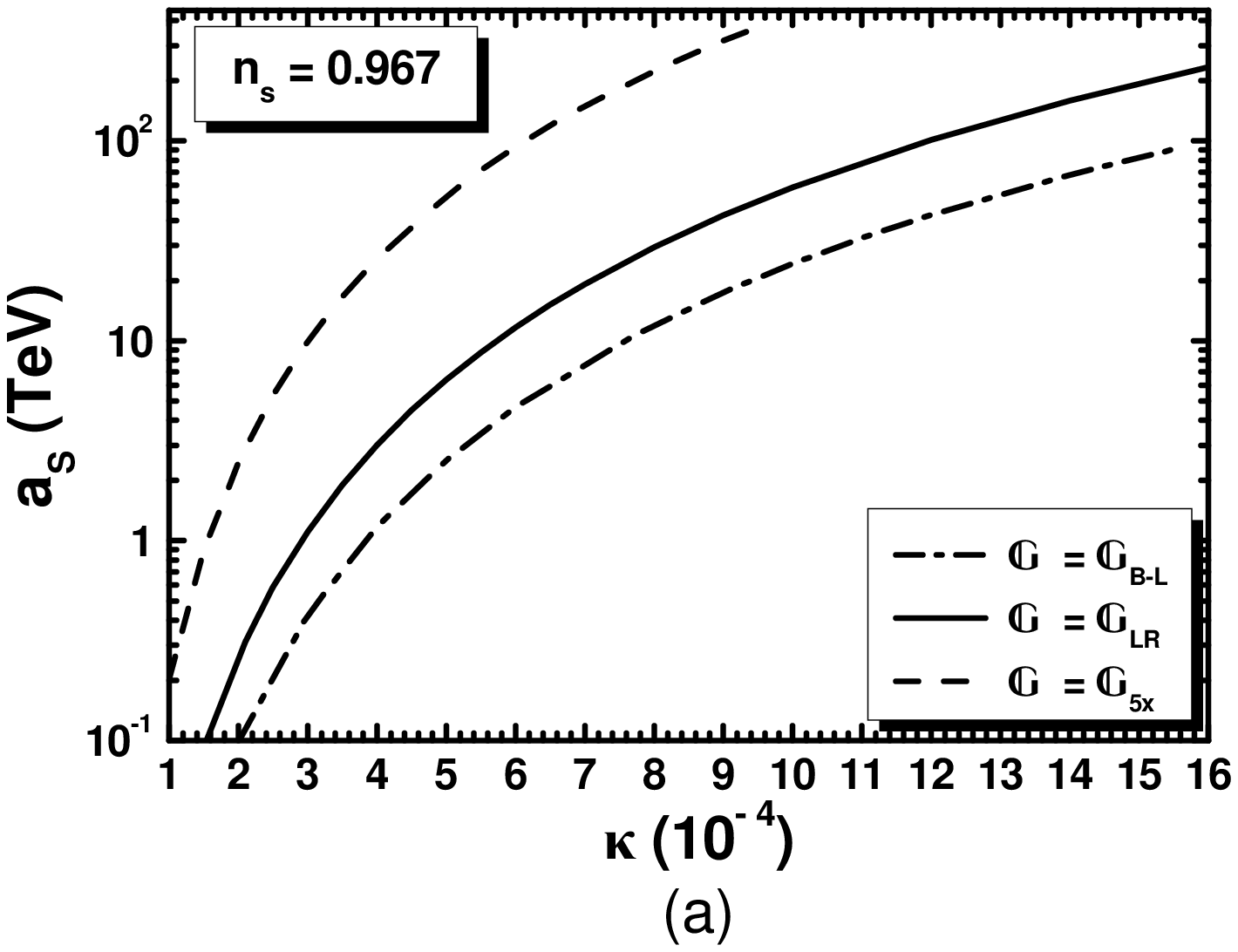,height=3.3in,angle=-90} \hspace*{-.2cm}
\epsfig{file=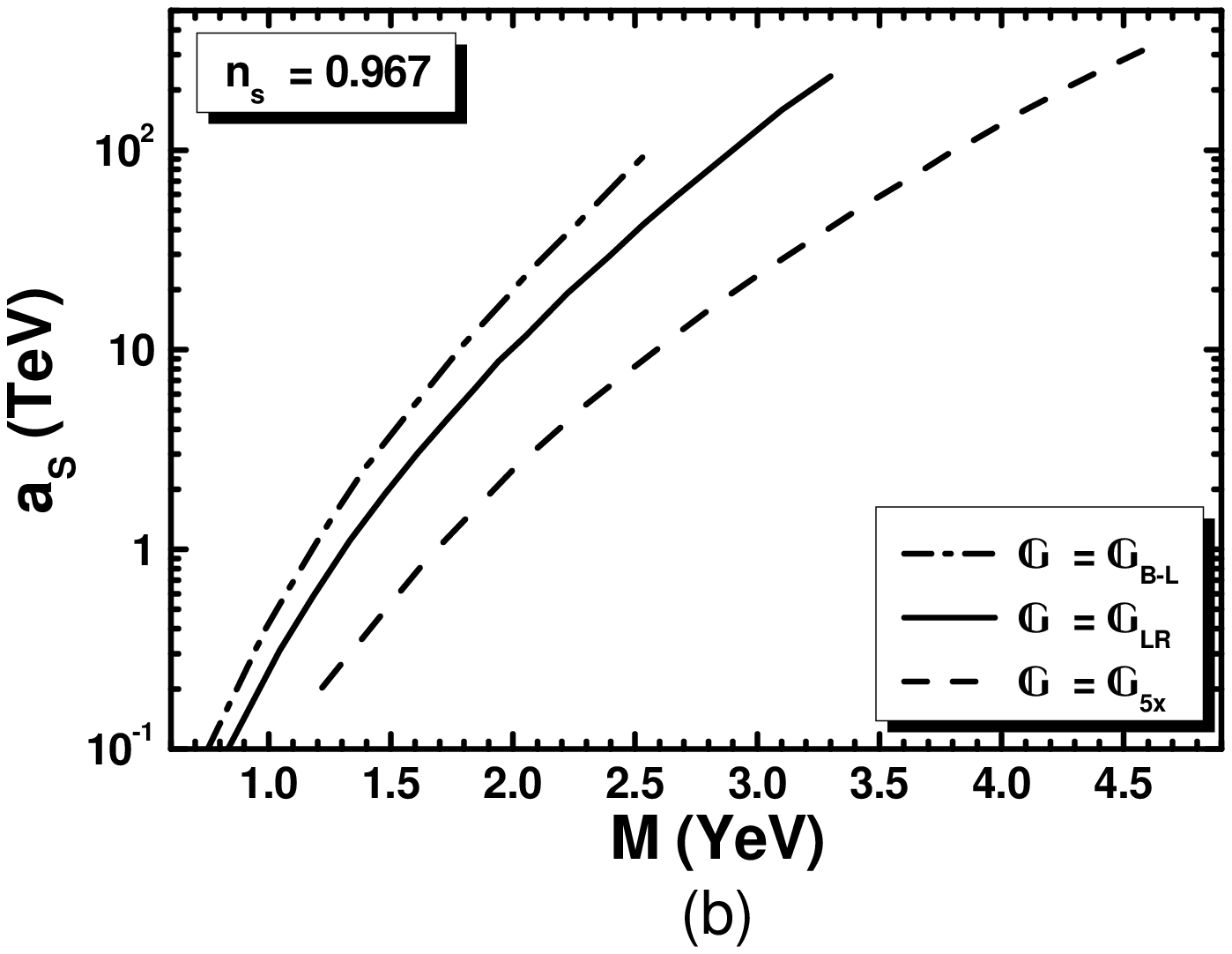,height=3.3in,angle=-90} \hfill
\end{minipage}
\hfill \caption{\sl\small Values of $\aS$ allowed by Eqs.~(3.7),
(3.8) and (3.10) versus $\kp$ ({\sf\ftn a}) and $M$ ({\sf\ftn b})
for various $\Ggut$'s and fixed $\ns=0.967$.}\label{fig2}
\end{figure}%(\ref{Ntot}), (\ref{Prob}) and (\ref{nswmap})

We also display in \Fref{fig2} the contours which are allowed by
Eqs.~(\ref{Nhi}) and (\ref{Prob}) in the $\kappa-\aS$ and $M-\aS$
planes taking $\ns=0.967$ and $\Ggut=\Gbl$ (dot-dashed line),
$\Ggut=\Glr$ (solid line) and $\Ggut=\Gfl$ (dashed line). The
various lines terminate at $\kp$ values close to $10^{-3}$ beyond
which no observationally acceptable inflationary solutions are
possible. From the plotted curves we notice that the required
$\aS$'s increase with $\Nr$. In particular, we end up with the
following ranges:
\beqs\bel &0.7\lesssim
{M/{\rm YeV}}\lesssim2.56~~\mbox{and}~~0.1\lesssim{\aS/\TeV}\lesssim100~~\mbox{for}~~\Ggut=\Gbl, \label{res1}  \\
&0.82\lesssim{M/{\rm
YeV}}\lesssim3.7~~\mbox{and}~~0.09\lesssim{\aS/\TeV}\lesssim234~~\mbox{for}~~\Ggut=\Glr,
\label{res2}\\& 1.22\lesssim {M/{\rm
YeV}}\lesssim4.77~~\mbox{and}~~0.2\lesssim{\aS/\TeV}\lesssim460~~\mbox{for}~~\Ggut=\Gfl
.\label{res3}\end{align} \eeqs\\ [-0.4cm]
The lower bounds of these inequalities  are expected to be
displaced to slightly larger values due to the BBN restrictions --
see \Sref{reh} below -- which are not considered here for the
shake of generality.

\section{Post-Inflationary Consequences}\label{pres}

We here explore the post-inflationary implications of our model
related to the present vacuum -- in \Sref{des} --, the
commencement of the radiation dominated era -- see \Sref{reh} --,
the SUSY scale -- see \Sref{susy} -- and the observational
consequences of CSs -- see \Sref{cssec}.

\subsection{\boldmath SUSY and $\Ggut$ Breaking -- Dark
Energy}\label{des}

The vacuum of our model is determined by minimizing the F--term
(tree level) SUGRA scalar potential $V_{\rm F}$ derived
\cite{asfhi} from $W$ in Eq.~(\ref{Who}) and $K$ in \Eref{Kho}.
This lies along the D-flat direction $|\bar\Phi|=|\Phi|$. Indeed,
as verified numerically \cite{asfhi}, $\vf$ is minimized at the
\Ggut-breaking vacuum
\beq \left|\vev{\Phi}\right|=\left|\vev{\bar\Phi}\right|=M.
\label{vevs} \eeq
Substituting \Eref{Zpara} in $\vf$ and minimizing it w.r.t the
various directions we arrive at the results
\beq \label{vevsg}\vev{\sg}\simeq2^{(1-\nu)/2}\lf m-\ld\mP\rg\
z^\nu/\mP^{\nu}~~\mbox{and}~~\vev{z}=2\sqrt{2/3}|\nu|\mP,\eeq
which yield the constant potential energy density
\beq \vev{\vf}\simeq\lf\frac{16\nu^{4}}{9}\rg^\nu \lf\frac{\ld
M^2-m\mP}{\kp\mP^2}\rg^2\om^N\mP^2\lf\ld\mP-m\rg^2~~
\mbox{with}~~\om\simeq\frac{2(3-2\nu)}{3}.\label{vcc}\eeq
Tuning $\ld$ to a value $\ld\sim m/\mP\simeq10^{-12}$ we may wish
identify $\vev{\vf}$ with the DE energy density, i.e.,
\beq \label{omde} \vev{\vf}=\Omega_\Lambda\rho_{\rm
c}=7.3\cdot10^{-121}\mP^4,\eeq
where the density parameter of DE and the current critical energy
density of the universe are respectively given by \cite{plcp}
\beq \label{rhoc} \Omega_\Lambda=0.6889~~\mbox{and}~~\rho_{\rm
c}=2.31\cdot10^{-120}h^2\mP^4~~\mbox{with}~~h=0.6766.\eeq
Therefore, we obtain a post-inflationary dS vacuum with
$\vev{\sg}\simeq0$ which explains the notorious DE problem. The
particle spectrum of the theory at the vacuum in \eqs{vevs}{vevsg}
includes the gravitino ($\Gr$) with mass \cite{susyr}
\beqs\beq \label{mgr} \mgro=\vev{e^{{\khh}/{2\mP^2}}W_{\rm
H}/\mP^2}\simeq 2^{\nu}3^{-\nu/2} |\nu|^{\nu}m\omega^{N/2},\eeq
the IS with mass
\beq \msn=e^{{\khh}/{2\mP^2}}\sqrt{2}\lf\kp^2
M^2+(4\nu^{2}/3)^\nu(1+4M^2/\mP^2)m^2\rg^{1/2},\label{msn}\eeq
which acquires some contribution from HS, the (canonically
normalized) sgoldstino (or $R$ saxion) and the pseudo-sgoldstino
(or $R$ axion) with respective masses
\beq \mz\simeq{3\om}\mgro/{2\nu} ~~\mbox{and}~~
\mth\simeq12k\om^{3/2}\mgro. \label{mzth}\eeq\eeqs
Numerical values for the masses above for the inputs of
\Tref{tab3} are given in the Table of \fref{fig4} for the various
$\Ggut$ in Eqs.~(\ref{gbl}) -- (\ref{gfl}). Shown is also there
the dimensionless SUGRA potential $\vf/m^2\mP^2$ as a function of
$z$ and $\sg$ for the inputs of Table 3 and $\Ggut=\Glr$. The
location of the dS vacuum in \Erefs{vevs}{vevsg} is depicted by a
thick point.

At the vacuum of \Eref{vevsg} $\dK$ in \Eref{dK} gives rise to a
non-vanishing $\mu$ term in the superpotential whereas the
contributions $W_{\rm Y}$ and $|Y_\al|^2$ of $W$ and $K$ in
\eqs{Who}{Kho} lead to a common soft SUSY-breaking mass parameter
$\mss$. Namely, we obtain \cite{susyr,asfhi}
\beq W\ni \mu  H_u H_d ~~\mbox{with}~~|\mu|=
\lm\lf{4\nu^2}/{3}\rg^\nu(5-4\nu)\mgr~~\mbox{and}~~
\mss=\mgr.\label{mssi}\eeq
The latter quantity indicatively represents the mass level of the
SUSY partners.

\begin{figure}[t]\vspace*{-.35in}
\begin{minipage}{75mm}
\vspace*{-4.cm}\epsfig{file=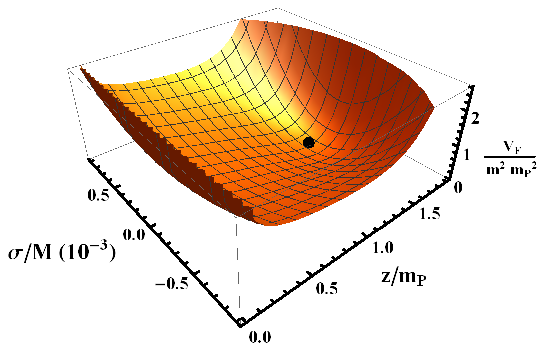,width=8.3cm,angle=-0}\end{minipage}
\begin{minipage}{75mm} \begin{center}\renewcommand{\arraystretch}{1.3}
\vspace*{.7in}\hspace*{0.4in}{\begin{tabular}{|c||c|c|c|c|}\hline
\Ggut&$\msn/$&$\mz/$&$\mth/$&$\mgr/$\\
&$\ZeV$&$\PeV$&$\PeV$&$\PeV$\\\hline\hline
\Gbl&$1.8$&$1.3$&$0.8$&$0.9$\\\hline
\Glr&$2.4$&$2.9$&$1.8$&$2$\\\hline
\Gfl&$4.5$&$16$&$10$&$11.2$\\\hline
\end{tabular}}\eec
\end{minipage}
\vspace*{.40in}\caption{\sl Dimensionless SUGRA potential
$\vf/m^2\mP^2$ as a function of $z$ and $\sg$ for the inputs of
Table 3 and $\Ggut=\Glr$. Shown is also the location of the dS
vacuum in Eq.~(4.1) and (4.2), depicted by a thick point. The
particle mass spectra for the input of Table~3 are also given in
the Table for the various $\Ggut$'s considered -- recall that
$1~\ZeV=10^{12}~\GeV=10^6~\PeV$.} \label{fig4}\end{figure}
\renewcommand{\arraystretch}{1.}

\subsection{Reheating Stage}\label{reh}

After the termination of FHI, the IS and $z$ start to oscillate
about their minima in \eqs{vevs}{vevsg} and eventually decay. When
the Hubble rate becomes $H_{z\rm I}\sim\mz$, the $z$ condensate
starts to dominate the universal energy budget, since the initial
energy density of its oscillations $\rho_{z\rm I}$  is comparable
to the energy density of the universe $\rho_{z\rm It}$
\beq \rho_{z\rm I}\sim\mz^2\vev{z}^2~~\mbox{with}~~\vev{z}\sim\mP
~~\mbox{and}~~\rho_{z\rm It}=3\mP^2H_{z\rm
I}^2\simeq3\mP^2\mz^2.\eeq
Due to the weakness of the gravitational interactions which govern
the $z$ decay, the reheating temperature
\beq \label{Trh} \Trh= \left({72/5\pi^2g_{\rm
rh*}}\right)^{1/4}\Gsn^{1/2}\mP^{1/2}, ~~\mbox{with}~~g_{\rm
rh*}\simeq10.75-100\eeq
the effective number of the relativistic degrees of freedom,  is
rather low -- in accordance with our expectations related to the
cosmic moduli problem \cite{baerh}. Indeed, the total decay width
of the (canonically normalized) sgoldstino
\beq\Gsn\simeq\Gth+\Gh,\label{Gol}\eeq
where the individual decay widths are found to be
\beqs \beq \Gth\simeq\frac{\ld_\theta^2\mz^3}{32\pi
\mP^2}\sqrt{1-\frac{4\mth^2}{\mgr^2}}
~~\mbox{and}~~\Gh=\frac{2^{4\nu-1}}{3^{2\nu-1}}\lm^2\frac{\om^2}{4\pi}
\frac{\mz^3}{\mP^2}\nu^{4\nu},\label{Gth}\eeq\eeqs
with $\ld_\theta={\vev{z}}/{N}\mP=({4\nu-3})/{\sqrt{6}\nu}$. From
the expressions above we readily recognize that $\Gsn$ is roughly
proportional to $\mz^3/\mP^2$ as expected for any typical modulus
\cite{baerh}. Selecting $\nu>3/4$ we kinematically forbid the
decay of $\dzh$ into $\Gr$'s and so we avoid the possible late
decay of the produced $\Gr$ and the troubles with abundance of the
subsequently produced lightest SUSY particles.

The compatibility between theoretical and observational values of
light element abundances predicted by BBN entails \cite{nsref} a
lower bound on $\Trh$ as follows
\beq \Trh\geq4.1~\MeV~~\mbox{for}~~\br=1~~\mbox{and}~~
\Trh\geq2.1~\MeV~~\mbox{for}~~\br=10^{-3},\label{tns}\eeq
where $\br$ is the hadronic branching ratio and large
$\mz\sim0.1~\PeV$ is assumed.

\begin{figure}[!t]\vspace*{-.15in}
\hspace*{-.12in}
\begin{minipage}{8in}
\epsfig{file=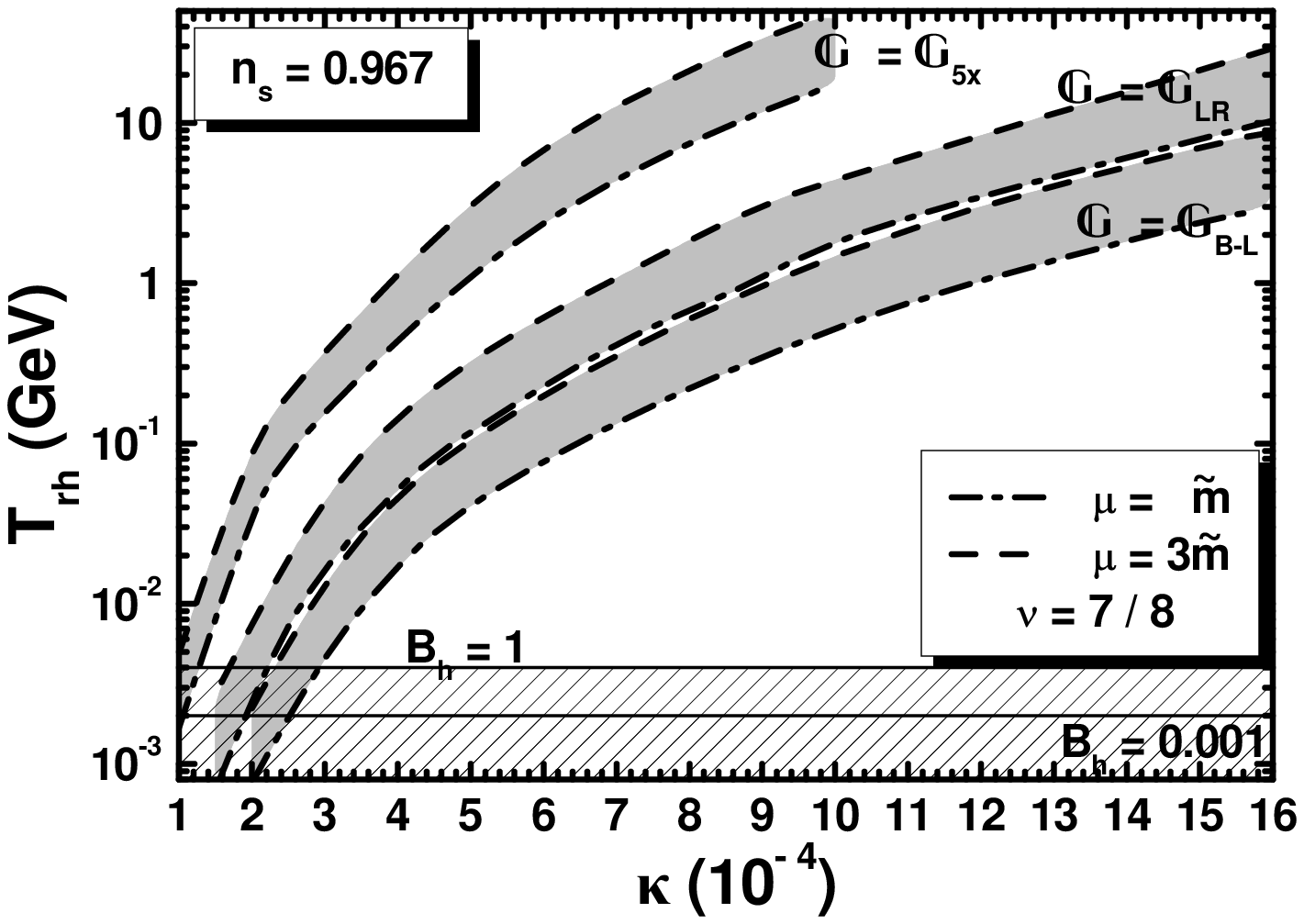,height=3.3in,angle=-90} \hspace*{-.2cm}
\epsfig{file=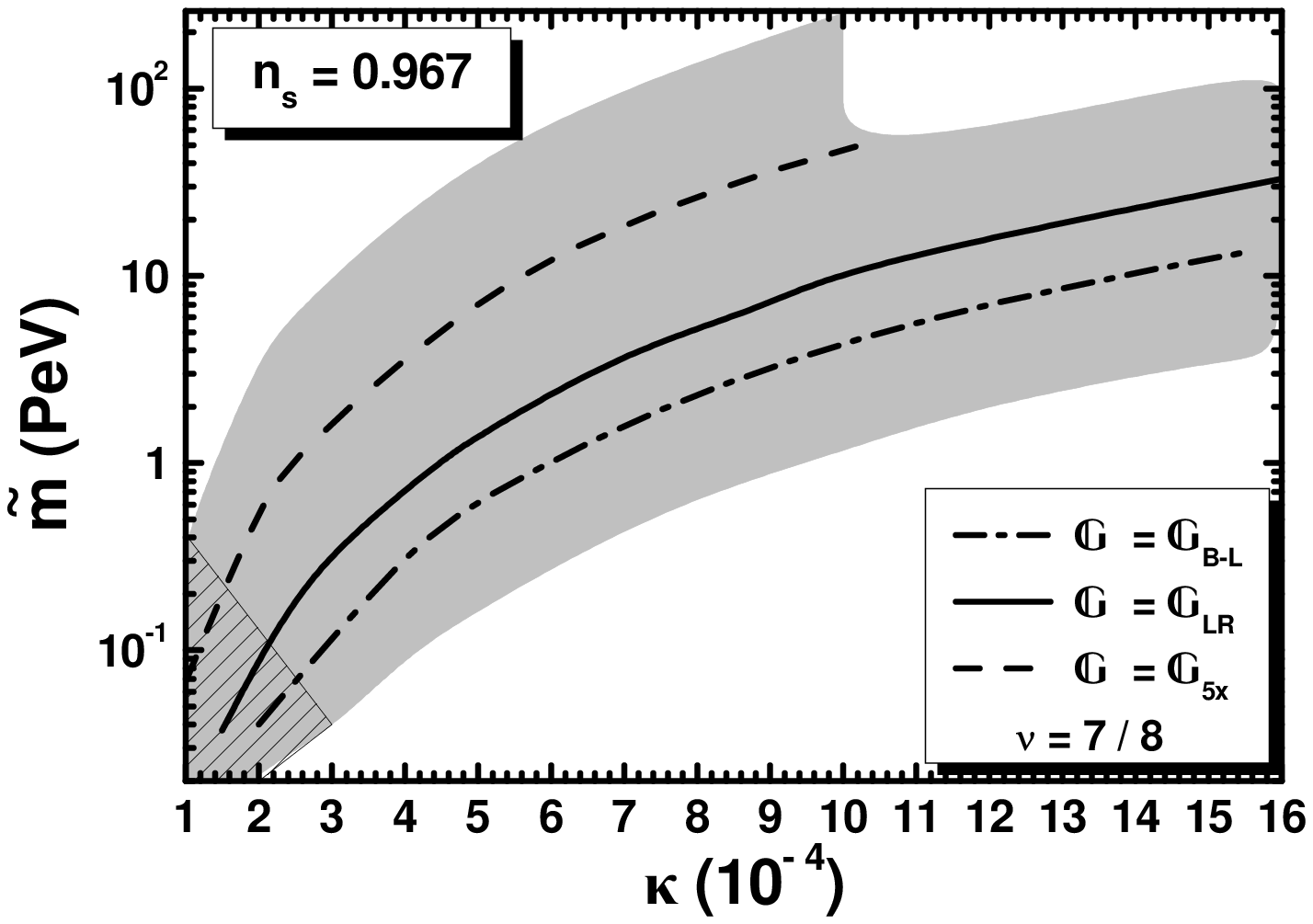,height=3.3in,angle=-90} \hfill
\end{minipage} \vspace*{-.47in}
\begin{flushleft}
\begin{tabular}[!h]{ll}
\hspace*{-.12in}
\begin{minipage}[t]{7.8cm}\caption[]{\sl Allowed strips in the $\kp-\Trh$ plane compatible
with the inflationary requirements in Sec.~3.3 for $\ns=0.967$,
and various $\Ggut$ indicated in the graph. We take $\nu=7/8$, and
$\mu=\mss$ (dot-dashed lines) or $\mu=3\mss$ (dashed lines). The
BBN lower bounds on $\Trh$ for hadronic branching ratios $\br=1$
and $0.001$ are also depicted by two thin
lines.}\label{fig5}\end{minipage}
&\begin{minipage}[t]{7.5cm}\caption[]{\sl Region in the $\kp-\mss$
plane allowed by the inflationary requirements in Sec.~3.3 for
$\ns=0.967$, $\mss\leq \mu\leq3\mss$, $1\leq\Nr\leq10$ and
$3/4<\nu<1$. The allowed contours for $\nu=7/8$ and specific
$\Ggut$ are also depicted. Hatched is the region excluded by BBN
for $\br=0.001$.}\label{fig6}
\end{minipage}
\end{tabular}
\end{flushleft}\vspace*{-.11in}
\end{figure}
%%%%%%%%%%%%%%%%%%%%%%%%%%%%%%%%%%%%%%%%$\lt=10^{-6}$ and

Taking $\kp$ and $\mz$ values allowed by the inflationary part of
our model we find the allowed regions in $\kp -\Trh$ plane,
displayed in \Fref{fig5}, for $\nu=7/8$ and the various $G$
considered here. The boundary curves of the allowed regions
correspond to $\mu=\mss$ or $\lm=0.65$ (dot-dashed line) and
$\mu=3\mss$ or $\lm=1.96$ (dashed line). The $|\mu|/\mss-\lm$
correspondence is determined via \Eref{mssi}. We see that there is
an ample parameter space consistent with the BBN bounds depicted
by two horizontal lines. The maximal values of $\Trh$ for the
selected $\nu$ are obtained for $\mu=3\mss$ and are estimated to
be
\beq T_{\rm
rh}^{\max}\simeq\begin{cases}14~\GeV~&\mbox{for}~~\Ggut=\Gbl,
\\33~\GeV~&\mbox{for}~~\Ggut=\Glr,
\\49~\GeV~&\mbox{for}~~\Ggut=\Gfl.\end{cases}
\label{trhb}\eeq
Obviously, reducing $\mu$ below $\mss$, the parameters $\lm$,
$\Gsn$, and so $\Trh$ decrease too and the slice cut by the BBN
bound increases. Therefore, our setting fits better with
high-scale SUSY \cite{strumia}.

\subsection{SUSY-Mass Scale}\label{susy}

Solving \Eref{aSn} w.r.t $m$ we obtain
\beq
m\simeq\lf\frac{\aS}{2^{1+\nu}(2-\nu)}\rg^{(2-\nu)/2}\lf\frac{3\Hhi^2}
{(1-\nu)\nu^2}\rg^{\nu/4}.\label{maS}\eeq
Taking into account \Eref{veviz} which results
$\vevi{z}/\mP\sim10^{-3}\ll1$ we estimate $m\sim10^3\aS$.
Inserting this result into \Eref{mgr} and then into the rightmost
relation of \Eref{mssi} we are able to gain information about
$\mss$ using as inputs the allowed $\aS$ values in
\Erefs{res1}{res3} -- which are compatible with the observational
constraints to FHI. Since $\aS\sim1~\TeV$ we expect $m\sim1~\PeV$,
and so $\mss\sim1~\PeV$ as well.

The qualitative arguments above are verified numerically in
\Fref{fig6} where we delineate the gray shaded region allowed by
the inflationary requirements of \Sref{fhi3} for $\ns=0.967$ by
varying $\nu$, $\mu$ and $\Nr$ within their possible respective
margins
\beq
0.75\lesssim\nu\lesssim1,~1\lesssim\mu/\mss\lesssim3~~\mbox{and}~~1\leq\Nr\leq10.
\eeq
Obviously the lower boundary curve of the displayed region is
obtained for $\Ggut=\Gbl$ and $\nu\simeq0.751$, whereas the upper
one corresponds to $\Ggut=\Gfl$ and $\nu\simeq0.99$. The hatched
region is ruled out by \Eref{tns}. All in all, we obtain the
predictions
\beq 1.2\lesssim \aS/\TeV\lesssim
460~~\mbox{and}~~0.09\lesssim\mss/\PeV\lesssim 253~~\mbox{with}~~
\label{msst}\eeq
and $T_{\rm rh}^{\max}\simeq71~\GeV, 139~\GeV$ and $163~\GeV$ for
$\Ggut=\Gbl, \Glr$ and $\Gfl$ respectively attained for
$\mu=3\mss$ and $\nu\simeq0.99$. Fixing $\nu=7/8$ and $\mu=\mss$,
we obtain a prediction for $\mss$ as a function of $\kp$ is
depicted  for the three $\Ggut$'s considered in our work. We use
the same type of lines as in \Fref{fig2}. Assuming also that  we
can determine the segments of these lines that can be excluded by
the BBN bound in \Eref{tns}. In all, we find that $\mss$ turns out
to be confined in the ranges
\beqs\bea & 0.34\lesssim\mss/\PeV\lesssim 13.6
~~\mbox{for}~~\Ggut=\Gbl,\label{mss1}\\
&0.21\lesssim\mss/\PeV\lesssim 32.9 ~~\mbox{for}~~\Ggut=\Glr,\label{mss2}\\
&0.58\lesssim\mss/\PeV\lesssim 46.8 ~~\mbox{for}~~\Ggut=\Gfl.
\label{mss10}\eea\eeqs

These results in conjunction with the necessity for $\mu\sim\mss$,
established in \Sref{reh} hint towards the \PeV-scale MSSM. Our
findings are compatible with the mass of the Higgs boson
discovered in LHC \cite{strumia} for degenerate sparticle spectrum
$1\leq\tan\beta\leq50$ and varying the stop mixing.

\subsection{Gravitational Waves from Cosmic Strings}\label{cssec}

When $\Ggut=\Gbl$, CSs may be produced after FHI with tension
\cite{mfhi}
\begin{equation} \label{mucs} \mcs \simeq
\frac12\lf\frac{M}{\mP}\rg^2\ecs(\rcs)~~\mbox{with}~~\ecs(\rcs)=\frac{2.4}{\ln(2/\rcs)}~~
\mbox{and}~~\rcs=\kappa^2/8g^2\leq10^{-2},\end{equation}where we
take into account that $(B-L)(\phc)=2$. Here $G=1/8\pi\mP^2$ is
the Newton gravitational constant and $g\simeq0.7$ is the gauge
coupling constant at a scale close to $M$. For the parameters in
\Eref{res1} we find
\beq 0.59\lesssim\mcs/10^{-8}\lesssim9.2.\label{rescs}\eeq

If the CSs are \emph{stable}, the corresponding parameter space is
totally allowed by the level of the CS contribution to the
observed anisotropies of CMB which is confined by \plk\
\cite{plcs0} in the range
\beq \mcs\lesssim 2.4\cdot 10^{-7}~~\mbox{at 95$\%$ c.l.}
\label{plcs} \eeq
On the other hand, the results of \Eref{rescs} are completely
excluded by the recent \emph{Pulsar Timing Array} ({\sf\ftn PTA})
bound which requires \cite{nano1}
\beq \mcs\lesssim 2\cdot 10^{-10}~~ \mbox{at 95$\%$ c.l.}
\label{ppta} \eeq

However, if these CSs are \emph{metastable} due to the embedding
of $\Gbl$ into a larger gauge group, whose spontaneous breaking to
$\Gbl$ produces monopoles -- see e.g. \cref{leont, nasri} --, an
explanation of the PTA data on the stochastic background of GWs is
possible for
\beq 0.9\lesssim
M/\YeV\lesssim2.56~~\mbox{and}~~3\lesssim\kp/10^{-4}\lesssim16.\label{rescs1}\eeq
Indeed, the obtained $\mcs$ values, through \Eref{mucs}, for the
values above lie within the range dictated by the interpretation
of the \emph{NANOGrav 15-years data} ({\sf\ftn \nano})
\cite{nano1}
\beq  4.3\cdot10^{-8}\lesssim  \mcs\lesssim 2.4\cdot
10^{-4}~~\mbox{for}~~8.2\gtrsim\sqrt{\rms}\gtrsim7.5~~ \mbox{at
95$\%$ c.l.}\label{kai} \eeq
Here the metastability factor $\rms$ is the ratio of the monopole
mass squared, $m_{\rm M}^2$, to $\mu_{\rm cs}$. Given that $m_{\rm
M}$ is related to the symmetry breaking scale of the GUT covering
$\Gbl$, the rightmost restriction in \Eref{kai} may constrain the
relevant scale close to the $M$ values in \Eref{rescs1}.

\section{Conclusions}\label{con}

We analyzed the implementation of FHI together with the SUSY
breaking within various GUTs in \eqss{gbl}{glr}{gfl}. We adopted
the super- and \Kaa\ potentials in \eqs{Who}{Kho} applying an
approximate $R$ symmetry. Our proposal offers the following
worth-mentioning achievements:

\begin{itemize}

\item Observationally acceptable FHI adjusting the tadpole
parameter, $\aS$, and the $\Ggut$-breaking scale $M$;

\item A prediction of the SUSY-mass scale, $\mss$, which turns out
to be of the order of \PeV;

\item Generation of the $\mu$ term of MSSM with $\mu \sim \mss$;

\item An interpretation of the DE problem without extensive
tuning;

\item Compatibility of $\Trh$ with BBN thanks to the considered
$\mu$ values;

\item An explanation of NG15 via the decay of metastable $B-L$ CSs
if $\Ggut=\Gbl$.

\end{itemize}

A complete cosmological picture of our framework could be achieved
if the issues of the correct baryon asymmetry and the
cold-dark-matter abundance of the universe are also addressed.

\def\ijmp#1#2#3{{\sl Int. Jour. Mod. Phys.}
{\bf #1},~#3~(#2)}
\def\plb#1#2#3{{\sl Phys. Lett. B }{\bf #1}, #3 (#2)}
\def\prl#1#2#3{{\sl Phys. Rev. Lett.}
{\bf #1},~#3~(#2)}
\def\rmp#1#2#3{{Rev. Mod. Phys.}
{\bf #1},~#3~(#2)}
\def\prep#1#2#3{{\sl Phys. Rep. }{\bf #1}, #3 (#2)}
\def\prd#1#2#3{{\sl Phys. Rev. D }{\bf #1}, #3 (#2)}
\def\npb#1#2#3{{\sl Nucl. Phys. }{\bf B#1}, #3 (#2)}
\def\npps#1#2#3{{Nucl. Phys. B (Proc. Sup.)}
{\bf #1},~#3~(#2)}
\def\mpl#1#2#3{{Mod. Phys. Lett.}
{\bf #1},~#3~(#2)}
\def\jetp#1#2#3{{JETP Lett. }{\bf #1}, #3 (#2)}
\def\app#1#2#3{{Acta Phys. Polon.}
{\bf #1},~#3~(#2)}
\def\ptp#1#2#3{{Prog. Theor. Phys.}
{\bf #1},~#3~(#2)}
\def\n#1#2#3{{Nature }{\bf #1},~#3~(#2)}
\def\apj#1#2#3{{Astrophys. J.}
{\bf #1},~#3~(#2)}
\def\mnras#1#2#3{{MNRAS }{\bf #1},~#3~(#2)}
\def\grg#1#2#3{{Gen. Rel. Grav.}
{\bf #1},~#3~(#2)}
\def\s#1#2#3{{Science }{\bf #1},~#3~(#2)}
\def\ibid#1#2#3{{\it ibid. }{\bf #1},~#3~(#2)}
\def\cpc#1#2#3{{Comput. Phys. Commun.}
{\bf #1},~#3~(#2)}
\def\astp#1#2#3{{Astropart. Phys.}
{\bf #1},~#3~(#2)}
\def\epjc#1#2#3{{Eur. Phys. J. C}
{\bf #1},~#3~(#2)}
\def\jhep#1#2#3{{\sl J. High Energy Phys.}
{\bf #1}, #3 (#2)}
\newcommand\jcap[3]{{\sl J.\ Cosmol.\ Astropart.\ Phys.\ }{\bf #1}, #3 (#2)}
\newcommand\njp[3]{{\sl New.\ J.\ Phys.\ }{\bf #1}, #3 (#2)}
\def\prdn#1#2#3#4{{\sl Phys. Rev. D }{\bf #1}, no. #4, #3 (#2)}
\def\jcapn#1#2#3#4{{\sl J. Cosmol. Astropart.
Phys. }{\bf #1}, no. #4, #3 (#2)}
\def\epjcn#1#2#3#4{{\sl Eur. Phys. J. C }{\bf #1}, no. #4, #3 (#2)}

\end{document}